# CHARACTERISTIC SPECIFIC PRIORITIZED DYNAMIC AVERAGE BURST ROUND ROBIN SCHEDULING FOR UNIPROCESSOR AND MULTIPROCESSOR ENVIRONMENT

Amar Ranjan Dash[1], Sandipta Kumar Sahu[2], Sanjay Kumar Samantra[2] and Sradhanjali Sabat[2]

[1]Department of Computer Science, Berhampur University, Berhampur, India
[2]Department of Computer Science, NIST, Berhampur, India

## ABSTRACT

*CPU scheduling is one of the most crucial operations performed by operating systems. Different conventional algorithms like FCFS, SJF, Priority, and RR (Round Robin) are available for CPU Scheduling. The effectiveness of Priority and Round Robin scheduling algorithm completely depends on selection of priority features of processes and on the choice of time quantum. In this paper a new CPU scheduling algorithm has been proposed, named as CSPDABRR (Characteristic specific Prioritized Dynamic Average Burst Round Robin), that uses seven priority features for calculating priority of processes and uses dynamic time quantum instead of static time quantum used in RR. The performance of the proposed algorithm is experimentally compared with traditional RR and Priority scheduling algorithm in both uni-processor and multi-processor environment. The results of our approach presented in this paper demonstrate improved performance in terms of average waiting time, average turnaround time, and optimal priority feature.*

## 1. INTRODUCTION

Operating systems are resource managers. The resources managed by Operating systems are hardware, storage units, input devices, output devices and data. Process scheduling is one of the functions performed by Operating systems. CPU scheduling is the method of selecting a process from the ready queue and allocating the CPU to it. Whenever CPU becomes idle, a waiting process from ready queue is selected and CPU is allocated to that. The performance of the scheduling algorithm mainly depends on CPU utilization, throughput, turnaround time, waiting time, response time, and context switch.

Conventionally four CPU scheduling techniques were there viz. FCFS, SJF, Priority, and RR. In FCFS, the process that requests the CPU first is allocated to the CPU first. In SJF, the CPU is allocated to the process with smallest burst time. In priority scheduling algorithm a priority is associated with each process, and the CPU is allocated to the process with the highest priority. In RR a small unit of time is used which is called Time Quantum or Time slice. The CPU scheduler goes around the Ready Queue allocating the CPU to each process for a time interval up to 1 time

DOI : 10.5121/ijcsea.2015.5501      1

quantum. If a process's CPU burst exceeds 1 time quantum, that process is pre-empted and is put back in the ready queue.

Different CPU scheduling algorithms described by Abraham Silberschatz et al. [1], viz. FCFS (First Come First Served), SJF (Shortest Job First), Priority and RR (Round Robin). Neetu Goel et al. [2] make a comparative analysis of CPU scheduling algorithms with the concept of schedulers. Jayashree S. Somani et al. [3] also make a similar analysis but with their characteristics and applications.

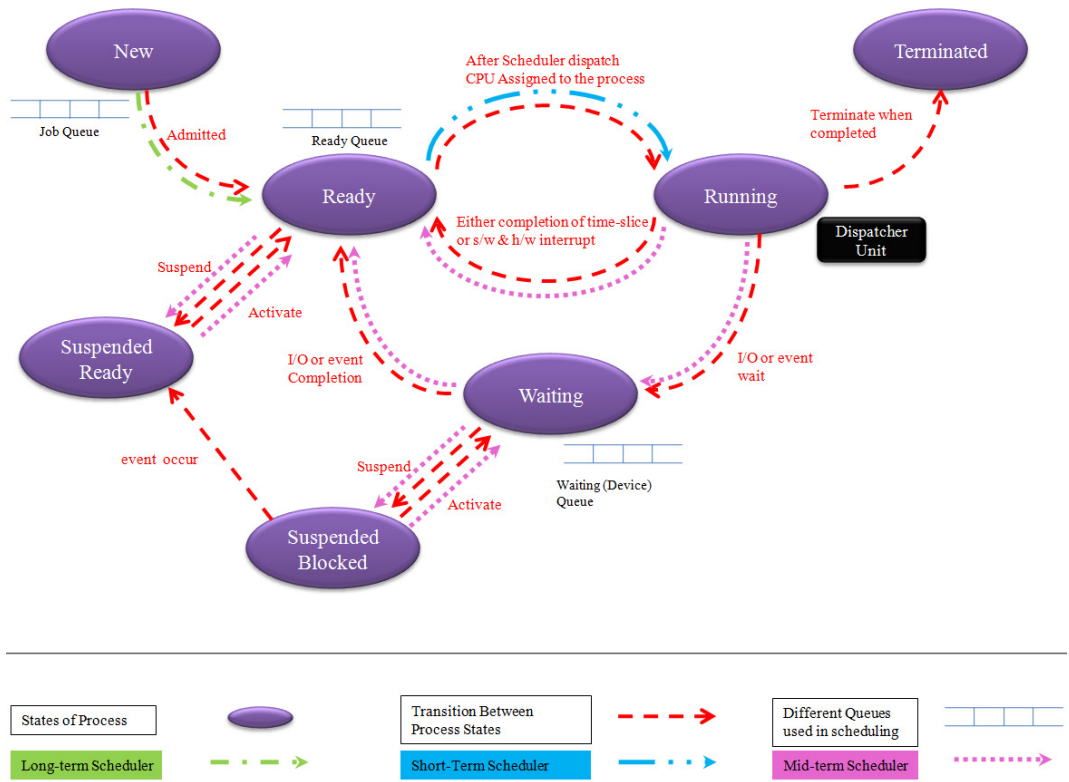

Figure 1: State Transition diagram of processes with different queues and schedulers.

Turnaround time is the time interval from the submission time of a process to the completion time of a process. Waiting time is the sum of periods spent waiting in the ready queue. The time from the submission of a process until the first response is called Response time. The CPU utilization is the percentage of time CPU remains busy. The number of processes completed per unit time is called Throughput. Context switch is the process of swap-out the pre-executed process from CPU and swap-in a new process to CPU. A scheduling algorithm can be optimized by minimizing response time, waiting time and turnaround time and by maximizing CPU utilization, throughput. Recently researchers try to improve the performance of Round Robin scheduling algorithm by manipulating the time slice. Rami J. Matarneh [4] designed an algorithm Self Adjustment Round Robin (SARR), in which after each cycle the median of burst time of the processes is calculated and used as time quantum. Abbas Noon et al. [5] develop an algorithm in which they calculate mean of burst time of all processes to use as time quantum. H.S.Behera et al. [6] also presents similar type of algorithm, but they rearrange the process during the next execution. It selects the process with lowest burst time, then process with highest burst time, then process with second lowest burst time, and so on.

Besides manipulating the time-slice some researchers combine the SJF and RR to improve the performance of RR scheduling algorithm. Among them Ajit Singh et al. [7] develop a scheduling algorithm in which in the first round they take a dynamic time quantum and then double the time quantum after every cycle. Manish Kumar Mishra et al. [8] developed a scheduling algorithm, which chooses the burst time of shortest process as new time quantum after each cycle. Rishi Verma [9] calculates the time quantum after every cycle by subtracting the minimum burst time from maximum burst time. Radhe Shyam et al. [10] also developed a scheduling algorithm in which they calculate the time quantum after every cycle by taking the square root of the multiplication of mean and the highest burst time. Amar Ranjan Dash et al. [11] developed an algorithm by arranging the processes in ascending order of burst time and by taking the average burst-time of the processes as time-slice to improve the performance of conventional Round Robin.

Bashir Alam et al. [12] develop Fuzzy Priority CPU Scheduling (FPCS) algorithm. They generate a fuzzy inference engine which gives a dynamic priority based on given static priority, Remaining Burst Time, and waiting time. H. S. Behera et al. [13] develop an Improved Fuzzy Based CPU Scheduling (IFCS) algorithm. They first calculate fuzzy membership value of priority ($\mu_p$), burst time ($\mu_b$), response ration ($\mu_h$) of individual processes, and then arrange the processes according to their membership value, which is the maximum value among $\mu_p$, $\mu_b$, $\mu_h$.

M. Ramakrishna et al. [14] and Ishwari Singh Rajput et al. [15] integrate the concept of priority scheduling with RR scheduling to optimize the Round Robin scheduling. They only consider shortness of the processes as priority component. H.S.Behera et al. in [16] also develop the same type of algorithm but they use the weighted mean of processes ($TQ_{wm}$) and the root mean square ($TQ_{rms}$). If the burst time of processes is less than the average burst time, then use $TQ_{wm}$ as Time Quantum. Otherwise use the addition of $TQ_{wm}$ and $TQ_{rms}$ as the Time Quantum. H.S.Behera et al. [17, 18] develop a prioritized round robin where they take shortness of the process and the number of context switch as priority component. They take a random time slice as original time slice. They add the priority components with original time slice and then calculate an intelligent time slice from that after each cycle. Zena Hussain Khalil et al. [19] modified the algorithm in [17] by adding the concept of Response ration into it.

H.S.Behera et al. [20] develop a round robin algorithm for two processor based real time system. They consider that one processor deal with CPU intensive process and another processor deal with I/O intensive process. They consider Gantt chart for each. But they don't include the priority component of processes in the algorithm. Previously researchers work on different scheduling algorithms, but in Uniprocessor environment. In this paper we work on a prioritized round robin technique for both uniprocessor and multiprocessor environment.

The rest of the paper is organized as follows: Section 2 presents a comparative analysis of conventional scheduling algorithm. Section 3 presents the proposed algorithm. In section 4, we make a comparative analysis of round robin, priority and our proposed algorithm experimentally with six test cases both in uniprocessor and multiprocessor environment. In section 5 we analyze the results obtained from our analysis. Section 6 provides the concluding remarks.

## 2. CONVENTIONAL SCHEDULING ALGORITHMS

Four conventional CPU scheduling algorithms are there FCFS, SJF, RR and Priority. FCFS scheduling algorithm arranges the process as per their arrival time. If more than one process arrives at the same time, FCFS scheduling algorithm unable to decide the sequence of processes in ready queue. SJF scheduling algorithm arranges the processes according to the ascending order of their burst time. But when more than one process arrive with same burst time, then SJF scheduling algorithm unable to decide the process sequence. Both FCFS and SJF scheduling algorithms are unable to optimize average turnaround time and average waiting time.

Among all conventional scheduling algorithms, the Round Robin scheduling algorithm provides optimized performance metics. To best of our knowledge, among all derived algorithms the DABRR [11] scheduling algorithm provides most optimized turnaround time and waiting time. The performance of round robin scheduling algorithm depends on its time quantum. But Round Robin scheduling algorithm is unable to differentiate process according to their priority and requirement. As an example if one system process and user process arrives, the system process should be given more preferences than user process. RR scheduling algorithm is unable to recognize these types of comparative solutions. Priority scheduling is algorithm able to classify the processes according to their priority. But it is unable to provide the optimized turnaround and waiting time. A major problem with priority scheduling is starvation. In this scheduling some low priority processes wait indefinitely to get allocated to the CPU.

In a multiprocessor environment, processes are provided to all processors in a balanced manner. Due to presence of multiple processors the load is balanced among them. The turn-around time and waiting time also gets decreased. But the problem arises during the selection of a particular processor for a process, load balancing, and resource management.

As per the above discussed criterion, the feature of Round Robin is required to improve the performance of processor. Similarly the feature of priority scheduling algorithm is required to avoid deadlock, etc. so the feature of both algorithms is important and unavoidable. So a new concept arises where the features of both the algorithms can be combined into one scheduling algorithm. In our proposed algorithm we consider all priority constraints of a process. We also integrate some features of DABRR in our algorithm. Our proposed algorithm is implemented considering both uniprocessor and multiprocessor environment.

## 3. OUR PROPOSAL

In our proposed algorithm we integrate the features of priority scheduling and round robin scheduling algorithm. Features of round robin can be improved by choosing a better time quantum for better performance metics. Similarly, the features of priority scheduling can be improved by selecting best method for choosing the priority of processes, for well arrangement of processes. At first for priority we include seven priority feature points.

A Priority Function Point is a value that is assigned to a particular process for its characteristics. PFP is the sum of values that are assigned to a particular process for particular characteristics index $PFP_i$. The priority value assigned has key value that is used further in CSPDABRR Scheduling .The PFP is automatically detected by the Operating System when the processes are in ready-queue. The Compiler analyzes the internal codes and verifies the criteria of each process for assigning PFP to it. Some process features used in our proposed algorithms are given below according to ascending order of characteristics index:

- $PFP_1$:

  - System Process: If the code is System Process that is completely originated by a system call, the process is termed as System Process and the $PFP_1$ is set to "1". Consider a process

"Services.exe" in windows Operating System which is completely System oriented process which runs automatically with the start of Operating System.
- ➢ User Process: If the process contains Kernel Calls rapidly but originated by User function calls then the process is termed as User Process and the $PFP_1$ is set to 2. Consider a process "Explorer.exe" in windows Operating System which is incited by login users which completely operates with system calls.

- ▪ $PFP_2$:

Types of process interrupts: Every Electronic System supports interrupts for support of multi tasking and acceptance of hardware interrupts to reliability and efficiency we can categorize the processes into two parts as follows

- ➢ Processes with hardware interrupts: Hardware interrupts are the interrupts that have been called directly by hardware units for I/O read, Memory Read, I/O write, Memory write the processes with max no. of any hardware command are considered as hardware interrupts. These processes has a high $PFP_2$ value as compared to software interrupts and the value is set to "3".consider a function "EvtInterruptIsr" for event handling in window operating system.
- ➢ Processes with software interrupts: Processes with software interrupt can be calculated by compile time. So they can be considered as criteria as they take a major part in program execution. If the no of software interrupt is count to be more than that of I/O calls in process then they are categorized under software interrupt list. The $PFP_2$ value is set to "2".In windows Operating system some software interrupts are CLI (Clear Interrupts), STI (SetInterrupts) and POPF (PopFlags).
- ➢ Process without any interrupts: the PFP2 value of processes without interrupts is set to "1". Like "Garbage collector" process runs periodically without any interrupt.

- ▪ $PFP_3$:

Execution Time Of task: Every process consists of codes and the execution time of the process can be calculated easily by compile time. Depending on bites of instruction, no of clock cycle and instruction length of processes can easily provide the information of execution time of process. If we consider the execution time of any tasks for determination of $PFP_3$ values we can classify tasks into three basic categories as follows:

- ➢ Anonymous Execution time: While compile time, if the process has very found that the instruction has not any time bound from its originating time then the task is categorized under Anonymous Execution time process and has a very High $PFP_3$ value and set to "3". NTLDR loader process in windows Operating System for booting.
- ➢ Medium calculable Execution time: While compile time, if any process execution time is calculated but can be extended due to specific regions then they are categorized under Medium calculable Execution time and the $PFP_3$ value is set to "2".WMIC.exe for information collection for data transfer.
- ➢ Real time Execution time: While compile time, if any process execution time is found dedicated to the process by the process originating function then the processes is categorized under Real time Execution time and the $PFP_3$ value is set to "1".Consider a function 'DWORD WINAPI SleepEx(_In_ DWORD dwMilliseconds,_In_ BOOL bAlertable);'responsible for sleep while battery is critically low.

- $PFP_4$:

Percentage of process completed: as per the percentage of process competed, at the time of arrival into the processor, we can classify the process into three categories. If the percentage of process completed is within the range of 67% to 100%, then the PFP4 is set to "1". When the percentage of process completed is within the range of 34% to 66% then the PFP4 is set to "2". When percentage of process completed is within the range of 0% to 33% then the PFP4 is set to "3".

- $PFP_5$:

Based on scheduling we classified the process into two type. If the process is half scheduled or organized then the $PFP_5$ of the process is set to "2". If the process is fully scheduled or organized then the $PFP_5$ of the process is set to "1".

- $PFP_6$:

Types of dependency: Every process in computer system has dependencies that is either hardware or software dependences. So the compiler while compile time can detect the degree of hardware or software dependences .So we can consider this criterion for determination of $PFP_6$ of any process. They can be categorized as follows:

➢ If the process has both hardware and software dependencies then $PFP_6$ of that process is set to "4".
➢ If the process has some hardware dependencies then $PFP_6$ of that process is set to "3".
➢ If the process has some software dependencies then $PFP_6$ of that process is set to "2".
➢ If the process does not have any dependencies then $PFP_6$ of that process is set to "1".

- PFP7:

Shortness component: if the burst time of the process is less than or equal to the mean of burst time than PFP7 is set to "1". Otherwise it considers being "2".

In the ready queue only we can calculate all seven features of a process. By adding these seven features we can calculate priority of a process. By viewing the priority value we can decide the priority level of process. As in normal task manager six priority levels are available. By studying from $PFP_1$ to $PFP_7$ the minimum value of priority can be 7 and the maximum value of priority can be 19. Lesser priority value indicates higher priority of process. As process with priority value 6 has higher priority than the process with priority value 9. The priority value can be classified into six priority level as the figure given below.

Table 1: Priority Values with respective priority level.

| Priority level | P1 | P2 | P3 | P4 | P5 | P6 |
|---|---|---|---|---|---|---|
| Priority Values | 7, 8 | 9, 10 | 11,12 | 13, 14 | 15,16 | 17,18,19 |

After finding priority value of each process, arrange processes in ascending order of priority values. If two processes have same priority values then arrange them according to the ascending order of burst time. Now find the average burst time of each process to set it as time quantum. Now if we are dealing with uniprocessor environment we can normally assign processor to the processes. But if we are dealing with multiprocessor environment then we again have to deal with load balancing. In multi-processing environment if more than one processor is free, then assign process to processors according to ascending order of processor. If we have same or less number of processes than the number of processor, then don't change process-processor bonding to stop unnecessary transfer of resources and information of one process from one processor to another.

### 3.1. CSPDABRR Algorithm

TQ     : Time Quantum  
RQ     : Ready Queue  
PQ     : Priority Queue  
TBT    : Total Burst Time  
$P_i$     : Process at $i^{th}$ index  
$PFP_j$  : $j^{th}$ Priority Feature Point  
n      : number of process in Ready Queue  
i      : used as index of ready queue or priority queue  
m      : number of processor in multi processing environment  
FLOOR: Mathematical function to found the largest number smaller than given float

```
CSPDABRR for multi-processor:
[1]   add all new process to the end of ready queue
[2]   n = number of processes in RQ
[3]   i=0
[4]   Repeat step 5 to 11 till i < n
[5]       j=1
[6]       Repeat step 7 to 9 till j <= 7
[7]           Calculate the PFPj for Pi
[8]           Priority of process Pi += PFPj
[9]           J++
[10]      PQ[i] = priority of the process Pi
[11]      i++
[12]  Arrange the processes with their respective priority,
      both in ready queue and priority queue, as per:
[13]      Ascending order of Priority. (lesser priority values
          indicate higher priority)
[14]      If two process has same priority then arrange them
          according to the ascending order of burst time
[15]  i=0, TBT=0
[16]  Repeat step 17 and 18 till i < n
[17]      TBT += burst time of process Pi
[18]      i++
[19]  TQ = FLOOR(TBT/n)
[20]  if(n <= m)
[21]      then don't change the process-processor bonding
[22]      if (burst time of Pi) <= TQ
[23]          Execute the process
[24]          Take the process out of RQ
[25]          n--
[26]      Else
[27]          Execute the process for a time interval up to 1
              TQ
[28]          Burst time of Pi = Burst time of Pi – TQ
[29]          Add the process to the end of ready queue for
              next round of execution
[30]  else
[31]      i = 0
[32]      Repeat from step 33 to 39 till i<n
[33]          Check which processor is free
[34]          if more than one processor is free
[35]              Assign process to processors according to
                  ascending order of processor
[36]          if (burst time of Pi) <= TQ
[37]              Execute the process
[38]              Take the process out of RQ
[39]              n--
[40]          Else
[41]              If(PQ[i] = = 6)
[42]                  Dedicate one processor to that process
[43]                  Take the process out of RQ
[44]                  n--
[45]              Else
[46]                  Execute the process for a time interval up
                      to 1 TQ
[47]                  Burst time of Pi = Burst time of Pi – TQ
[48]                  Add the process to the end of ready queue
                      for next round of execution
[49]              i++
[50]  If new process arrives
[51]      goto step 1
[52]  If RQ is not empty
[53]      goto step 3

CSPDABRR for uni-processor:
[1]   add all new process to the end of ready queue
[2]   n = number of processes in RQ
[3]   i=0
[4]   Repeat step 5 to 11 till i < n
[5]       j=1
[6]       Repeat step 7 to 9 till j <= 7
[7]           Calculate the PFPj for Pi
[8]           Priority of process Pi += PFPj
[9]           J++
[10]      PQ[i] = priority of the process Pi
[11]      i++
[12]  Arrange the processes with their respective priority,
      both in ready queue and priority queue, as per:
[13]      Ascending order of Priority. (lesser priority value
          indicate higher priority)
[14]      If two process has same priority then arrange them
          according to the ascending order of burst time
[15]  i=0, TBT=0
[16]  Repeat step 17 and 18 till i < n
[17]      TBT += burst time of process Pi
[18]      i++
[19]  TQ = FLOOR(TBT/n)
[20]  i = 0
[21]  Repeat from step 22 to 30 till i<n
[22]      if (burst time of Pi) <= TQ
[23]          Execute the process
[24]          Take the process out of RQ
[25]          n--
[26]      Else
[27]          Execute the process for a time interval up to TQ
[28]          Burst time of Pi = Burst time of Pi – TQ
[29]          Add the process to the end of ready queue for
              next round of execution
[30]      i++
[31]  If new process arrives
[32]      goto step 1
[33]  If RQ is not empty
[34]      goto step 3
```

Figure 2: CSPDABRR scheduling algorithm

### 3.2. Assumption

During analysis we have considered pre-emptive only. In each test case 5 processes are analyzed in both uniprocessor & multiprocessor environment. Corresponding burst time and arrival time of processes are known before execution. The context switch time of processes has been considered as zero. The time required for arranging the processes in ascending order of priority also considered as zero. We assume the static time quantum for round robin as 25.

## 4. EVALUATION

CASE1.    Processes without arrival time and with Ascending Burst time

  a. Uniprocessor Environment:

Table 2: Process of CASE-I.

| Process | P1 | P2 | P3 | P4 | P5 |
|---|---|---|---|---|---|
| A.T | 0 | 0 | 0 | 0 | 0 |
| B.T | 40 | 55 | 60 | 90 | 102 |
| Priority | 3 | 5 | 2 | 4 | 1 |

**RR(Round Robin):**

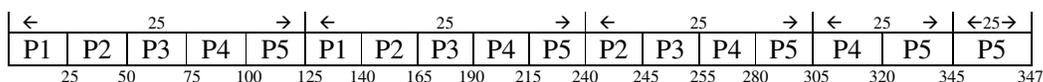

Average Turnaround time = (140+245+255+320+347)/5 = 1307/5 = 261.4
Average Waiting time = (100+190+195+230+245)/5 = 960/5 = 192

**Priority:**

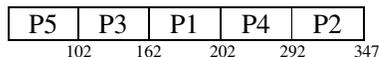

Average Turnaround time = (202+347+162+292+102)/5 = 1105/5 = 221
Average Waiting time = (162 + 292 + 102 + 202 + 0)/5 = 758/5 = 151.6

**CSPDABRR:**

Table 3: The scheduling table of CSPDABRR for CASE-I in uniprocessor environment.

| Round | 1st | | | | | 2nd | | 3rd |
|---|---|---|---|---|---|---|---|---|
| Process | P1 | P2 | P3 | P4 | P5 | P4 | P5 | P5 |
| B.T | 40 | 55 | 60 | 90 | 102 | 21 | 33 | 6 |
| A.T | 0 | 0 | 0 | 0 | 0 | 0 | 0 | 0 |
| PFP1 | 2 | 2 | 1 | 1 | 2 | 1 | 2 | 2 |
| PFP2 | 2 | 2 | 3 | 2 | 3 | 2 | 3 | 3 |
| PFP3 | 1 | 2 | 3 | 2 | 3 | 2 | 3 | 3 |
| PFP4 | 3 | 3 | 3 | 3 | 3 | 1 | 1 | 1 |
| PFP5 | 1 | 1 | 2 | 1 | 1 | 1 | 1 | 1 |
| PFP6 | 2 | 2 | 3 | 4 | 4 | 4 | 4 | 4 |
| PFP7 | 1 | 1 | 1 | 2 | 2 | 1 | 2 | 1 |
| Total | 12 | 13 | 16 | 15 | 18 | 12 | 16 | 15 |
| P.L | A.N | N | B.N | B.N | L | A.N | B.N | B.N |

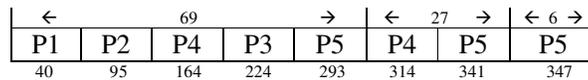

| P1 | P2 | P4 | P3 | P5 | P4 | P5 | P5 |
|---|---|---|---|---|---|---|---|
| 40 | 95 | 164 | 224 | 293 | 314 | 341 | 347 |

Average Turnaround time = (40+95+224+314+347)/5 = 1020/5 = 204
Average Waiting time = (0+40+164+224+245)/5 = 673/5 = 134.6

b. Multiprocessor Environment: (with 2 processors)

**RR(Round Robin):**

Average Turnaround time
= (65+120+130+165+182)/5
= 662/5 = 132.4
Average Waiting time
= (25+65+70+75+80)/5 = 315/5 = 63

**Priority:**
Average Turnaround time
= (100+157+60+190+102)/5
= 609/5 = 121.8
Average Waiting time
= (60+102+0+100+0)/5 = 262/5 = 52.4

**CSPDABRR:**
Average Turnaround time
= (40+55+115+136+211)/5
= 557/5 = 111.4
Average Waiting time
= (0+0+55+46+109)/5 = 210/5 = 42

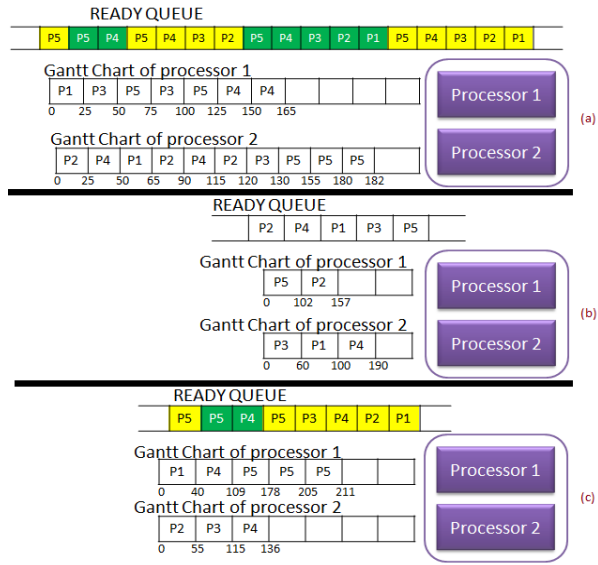

Figure 3: Pictorial representation of Round Robin, Priority, and CSPDABRR for CASE-I in Multiprocessor environment

CASE2. Processes without arrival time and with Descending Burst time

a. Uniprocessor Environment:

Table 4: Process of CASE-II.

| Process | P1 | P2 | P3 | P4 | P5 |
|---|---|---|---|---|---|
| A.T | 0 | 0 | 0 | 0 | 0 |
| B.T | 97 | 83 | 45 | 32 | 6 |
| Priority | 2 | 1 | 3 | 5 | 4 |

**RR(Round Robin):**

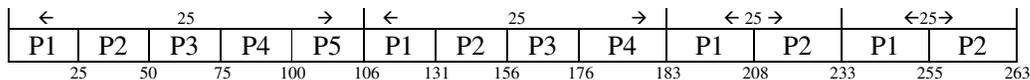

| P1 | P2 | P3 | P4 | P5 | P1 | P2 | P3 | P4 | P1 | P2 | P1 | P2 |
|---|---|---|---|---|---|---|---|---|---|---|---|---|
| 25 | 50 | 75 | 100 | 106 | 131 | 156 | 176 | 183 | 208 | 233 | 255 | 263 |

Average Turnaround time = (255+263+176+183+106)/5 = 983/5 = 196.6
Average Waiting time = (158+180+131+151+100)/5 = 720/5 = 144

**Priority:**

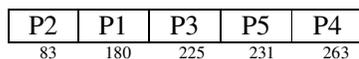

| P2 | P1 | P3 | P5 | P4 |
|---|---|---|---|---|
| 83 | 180 | 225 | 231 | 263 |

Average Turnaround time = (180+83+225+263+231)/5 = 982/5 = 196.4
Average Waiting time = (83 + 0 + 180 + 231 + 225)/5 = 719/5 = 143.8

**CSPDABRR:**

Table 5: The scheduling table of CSPDABRR for CASE-II in uniprocessor environment.

| Round | 1st | | | | | 2nd | | 3rd |
|---|---|---|---|---|---|---|---|---|
| Process | P1 | P2 | P3 | P4 | P5 | P1 | P2 | P1 |
| B.T | 97 | 83 | 45 | 32 | 6 | 45 | 31 | 7 |
| A.T | 0 | 0 | 0 | 0 | 0 | 0 | 0 | 0 |
| PFP1 | 2 | 1 | 2 | 2 | 1 | 2 | 1 | 2 |
| PFP2 | 2 | 3 | 2 | 1 | 1 | 2 | 3 | 2 |
| PFP3 | 3 | 2 | 1 | 1 | 2 | 3 | 2 | 3 |
| PFP4 | 3 | 3 | 3 | 3 | 3 | 2 | 2 | 1 |
| PFP5 | 1 | 2 | 1 | 2 | 1 | 1 | 2 | 1 |
| PFP6 | 3 | 4 | 2 | 1 | 1 | 3 | 4 | 3 |
| PFP7 | 2 | 2 | 1 | 1 | 1 | 2 | 1 | 1 |
| Total | 16 | 17 | 12 | 11 | 10 | 15 | 15 | 13 |
| P.L | B.N | L | A.N | A.N | H | B.N | B.N | N |

| ← | | 52 | | → | ← | 38 | → | ← 7 → |
|---|---|---|---|---|---|---|---|---|
| P5 | P4 | P3 | P1 | P2 | P2 | P1 | | P1 |
| 6 | 38 | 83 | 135 | 187 | 218 | 256 | | 263 |

Average Turnaround time = (263+218+83+38+6)/5 = 608/5 = 121.6
Average Waiting time = (166+135+38+6+0)/5 = 345/5 = 69

b. Multiprocessor Environment: (with 2 processors)

**RR(Round Robin):**

Average Turnaround time
= (135+128+95+88+56)/5 = 502/5
= 100.4
Average Waiting time
= (38+45+50+56+50)/5 = 239/5 = 47.8

**Priority:**

Average Turnaround time
= (97+83+128+103+135)/5
= 546/5 = 109.2
Average Waiting time
= (0+0+83+103+97)/5 = 283/5 = 56.6

**CSPDABRR:**

Average Turnaround time
= (129+134+51+32+6)/5
= 352/5 = 70.4
Average Waiting time
= (32+51+6+0+0)/5 = 89/5 = 17.8

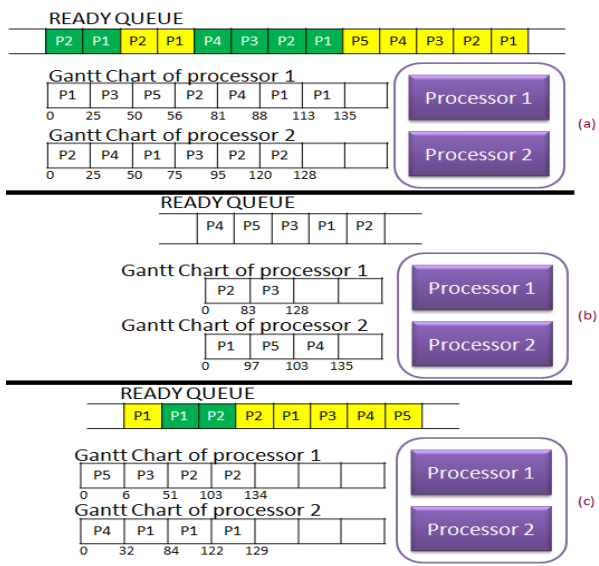

Figure 4: Pictorial representation of Round Robin, Priority, and CSPDABRR for CASE-II in Multiprocessor environment

CASE3.        Processes without arrival time and with random Burst time
   a. Uniprocessor Environment:

Table 6: Process of CASE-III.

| Process | P1 | P2 | P3 | P4 | P5 |
|---------|----|----|----|----|----|
| A.T     | 0  | 0  | 0  | 0  | 0  |
| B.T     | 12 | 32 | 6  | 54 | 83 |
| Priority| 2  | 1  | 4  | 3  | 5  |

**RR(Round Robin):**

| ← | 25 | → | ← | 25 | → | ← | 25 | → | ←25→ |
|---|----|---|---|----|---|---|----|---|------|
| P1 | P2 | P3 | P4 | P5 | P2 | P4 | P5 | P4 | P5 | P5 |
| 12 | 37 | 43 | 68 | 93 | 100 | 125 | 150 | 154 | 179 | 187 |

Average Turnaround time = (12+100+43+154+187)/5 = 496/5 = 99.2
Average Waiting time = (0+68+37+104+104)/5 = 313/5 = 62.6

**Priority:**

| P2 | P1 | P4 | P3 | P5 |
|----|----|----|----|----|
| 32 | 44 | 98 | 104 | 187 |

Average Turnaround time = (32+44+104+98+187)/5 = 465/5 = 93
Average Waiting time = (32 + 0 + 98 + 44 + 104)/5 = 278/5 = 55.6

**CSPDABRR:**

Table 7: The scheduling table of CSPDABRR for CASE-III in uniprocessor environment.

| Round   | 1st |    |    |    |    | 2nd |    | 3rd |
|---------|-----|----|----|----|----|-----|----|-----|
| Process | P1  | P2 | P3 | P4 | P5 | P4  | P5 | P5  |
| B.T     | 12  | 32 | 6  | 54 | 83 | 17  | 46 | 15  |
| A.T     | 0   | 0  | 0  | 0  | 0  | 0   | 0  | 0   |
| PFP1    | 1   | 2  | 1  | 1  | 2  | 1   | 2  | 2   |
| PFP2    | 1   | 2  | 2  | 3  | 2  | 3   | 2  | 2   |
| PFP3    | 2   | 3  | 1  | 2  | 3  | 2   | 3  | 3   |
| PFP4    | 3   | 3  | 3  | 3  | 3  | 1   | 2  | 1   |
| PFP5    | 1   | 2  | 1  | 2  | 2  | 2   | 2  | 2   |
| PFP6    | 1   | 2  | 2  | 3  | 4  | 3   | 4  | 4   |
| PFP7    | 1   | 1  | 1  | 2  | 2  | 1   | 2  | 1   |
| Total   | 10  | 15 | 11 | 16 | 18 | 13  | 17 | 15  |
| P.L     | H   | B.N| A.N| B.N| L  | N   | L  | B.N |

| ← | 37 | → | ← | 31 | → | ← 15 → |
|---|----|---|---|----|---|--------|
| P1 | P3 | P4 | P2 | P5 | P4 | P5 | P5 |
| 12 | 18 | 55 | 87 | 124 | 141 | 172 | 187 |

Average Turnaround time = (12+87+18+141+187)/5 = 445/5 = 89
Average Waiting time = (0+55+12+87+104)/5 = 258/5 = 51.6

   b. Multiprocessor Environment: (with 2 processors)

**RR(Round Robin):**

Average Turnaround time

= (12+50+18+79+108)/5

= 267/5 = 53.4
Average Waiting time
= (0+18+12+25+25)/5 = 80/5 = 16

**Priority:**

Average Turnaround time
= (12+32+38+66+121)/5
= 269/5 = 53.8
Average Waiting time
= (0+0+32+12+38)/5 = 82/5 = 16.4

**CSPDABRR:**

Average Turnaround time
= (12+44+6+61+126)/5
= 249/5 = 49.5
Average Waiting time
= (0+12+0+7+43)/5 = 62/5 = 12.4

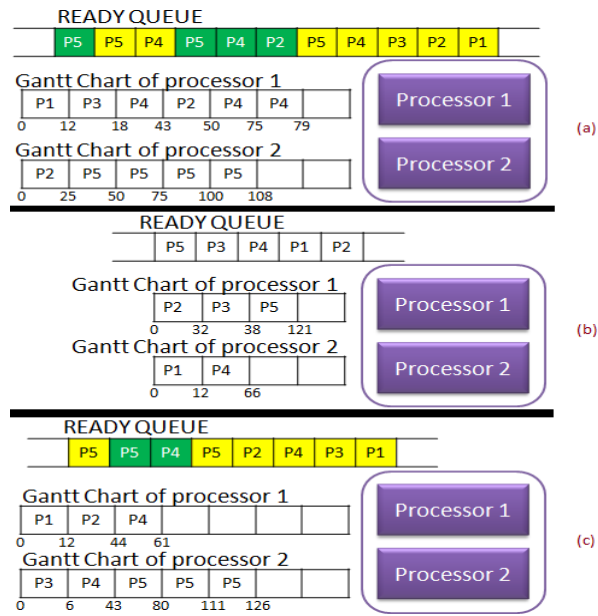

Figure 5: Pictorial representation of Round Robin, Priority, and CSPDABRR for CASE-III in Multiprocessor environment

CASE4.   Processes with arrival time and with Ascending Burst time
  a. Uniprocessor Environment:

Table 8: Process of CASE-IV.

| Process | P1 | P2 | P3 | P4 | P5 |
|---|---|---|---|---|---|
| A.T | 0 | 3 | 5 | 7 | 9 |
| B.T | 27 | 32 | 55 | 82 | 110 |
| Priority | 3 | 5 | 4 | 2 | 1 |

**RR(Round Robin):**

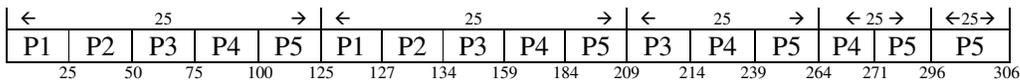

Average Turnaround time = (127 + (134-3) + (214-5) + (271-7) + (306-9))/5 = 1028/5 = 205.6
Average Waiting time = (100 + (102-3) + (159-5) + (189-7) + (196-9))/5 = 722/5 = 144.4

**Priority:**

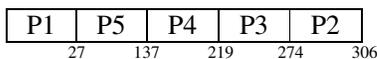

Average Turnaround time = (27 + (306-3) + (274-5) + (219-7) + (137-9))/5 = 939/5 = 187.8
Average Waiting time = (0 + (274-3) + (219-5) + (137-7) + (27-9))/5 = 633/5 = 126.6

International Journal of Computer Science, Engineering and Applications (IJCSEA) Vol.5, No.4/5, October 2015

**CSPDABRR:**

Table 9: The scheduling table of CSPDABRR for CASE-IV in uniprocessor environment.

| Round | 1st | 2nd | | | | 3rd | | 4th |
|---|---|---|---|---|---|---|---|---|
| Process | P1 | P2 | P3 | P4 | P5 | P4 | P5 | P5 |
| B.T | 27 | 32 | 55 | 82 | 110 | 13 | 41 | 14 |
| A.T | 0 | 3 | 5 | 7 | 9 | 7 | 9 | 9 |
| PFP1 | 1 | 2 | 2 | 1 | 1 | 1 | 1 | 1 |
| PFP2 | 3 | 2 | 3 | 2 | 3 | 2 | 3 | 3 |
| PFP3 | 3 | 3 | 2 | 3 | 1 | 3 | 1 | 1 |
| PFP4 | 3 | 3 | 3 | 3 | 3 | 1 | 2 | 1 |
| PFP5 | 1 | 1 | 2 | 2 | 1 | 2 | 1 | 1 |
| PFP6 | 4 | 2 | 3 | 4 | 4 | 4 | 4 | 4 |
| PFP7 | 1 | 1 | 1 | 2 | 2 | 1 | 2 | 1 |
| Total | 16 | 14 | 16 | 17 | 15 | 14 | 14 | 12 |
| P.L | B.N | N | B.N | L | B.N | N | N | A.N |

| ←27→ | ← | | 69 | → | ← | 27 | → | ←14→ |
|---|---|---|---|---|---|---|---|---|
| P1 | P2 | P5 | P3 | P4 | P4 | | P5 | P5 |
| 27 | 59 | 128 | 183 | 252 | 265 | | 292 | 306 |

Average Turnaround time = (27 + (59-3) + (183-5) + (265-7) + (306-9))/5 = 816/5 = 163.2
Average Waiting time = (0 + (27 – 3) + (128 - 5) + (183 - 7) + (196 - 9))/5 = 510/5 = 102

b. Multiprocessor Environment: (with 2 processors)

**RR(Round Robin):**

Average Turnaround time = (55 + (62-3) + (105-5) + (137-7) + (172-9))/5 = 507/5 = 101.4
Average Waiting time = (28 + (30 - 3) + (50 - 5) + (55 - 7) + (62-9))/5 = 201/5 = 40.2

**Priority:**

Average Turnaround time = (27 + (35-3) + (172-5) + (117-7) + (137-9))/5 = 464/5 = 92.8
Average Waiting time = (0 + (3-3) + (117-5) + (35-7) + (27-9))/5 = 158/5 = 31.6

**CSPDABRR:**

Table 10: The scheduling table of CSPDABRR for CASE-IV in multiprocessor environment.

| Round | 1st | 2nd | 3rd | | | 4th |
|---|---|---|---|---|---|---|
| Process | P1 | P2 | P3 | P4 | P5 | P5 |
| B.T | 27 | 32 | 55 | 82 | 110 | 28 |
| A.T | 0 | 3 | 5 | 7 | 9 | 9 |
| PFP1 | 1 | 2 | 2 | 1 | 1 | 1 |
| PFP2 | 3 | 2 | 3 | 2 | 3 | 3 |
| PFP3 | 3 | 3 | 2 | 3 | 1 | 1 |
| PFP4 | 3 | 3 | 3 | 3 | 3 | 1 |
| PFP5 | 1 | 1 | 2 | 2 | 1 | 1 |
| PFP6 | 4 | 2 | 3 | 4 | 4 | 4 |
| PFP7 | 1 | 1 | 1 | 1 | 2 | 1 |
| Total | 16 | 14 | 16 | 16 | 15 | 12 |
| P.L | B.N | N | B.N | B.N | B.N | A.N |

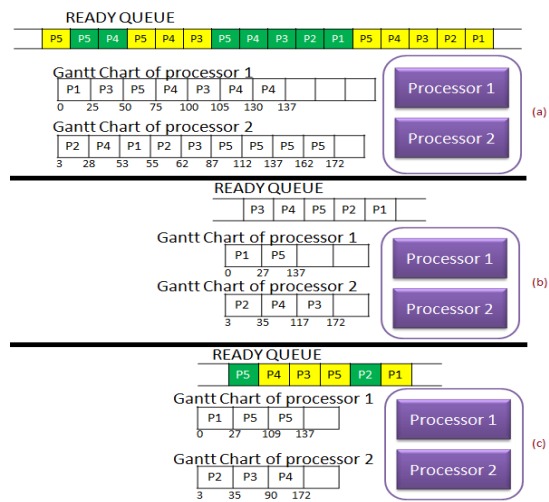

Figure 6: Pictorial representation of Round Robin, Priority, and CSPDABRR for CASE-IV in Multiprocessor environment

Average Turnaround time = (27 + (35-3) + (90-5) + (172-7) +(137-9))/5 = 437/5 = 87.4
Average Waiting time = (0 + (3-3) + (35-5) + (90-7) + (27-9))/5 = 131/5 = 26.2

CASE5.    Processes with arrival time and with Descending Burst time
  a.  Uniprocessor Environment:

Table 11: Process of CASE-V.

| Process | P1 | P2 | P3 | P4 | P5 |
|---------|----|----|----|----|----|
| A.T | 0 | 2 | 4 | 8 | 16 |
| B.T | 105 | 80 | 60 | 45 | 32 |
| Priority | 2 | 3 | 4 | 5 | 1 |

**RR(Round Robin):**

| ← | 25 | → | ← | 25 | → | ← | 25 | → | ← 25 → | ← 25→ |
|---|---|---|---|---|---|---|---|---|---|---|
| P1 | P2 | P3 | P4 | P5 | P1 | P2 | P3 | P4 | P5 | P1 | P2 | P3 | P1 | P2 | P1 |
| 25 | 50 | 75 | 100 | 125 | 150 | 175 | 200 | 220 | 227 | 252 | 277 | 287 | 312 | 317 | 322 |

Average Turnaround time = (322 + (317-2) + (287-4) + (220-8) + (227-16))/5 = 1343/5 = 268.6
Average Waiting time = (217 + (237-2) + (227-4) + (175-8) + (195-16))/5 = 1021/5 = 204.2

**Priority:**

| P1 | P5 | P2 | P3 | P4 |
|----|----|----|----|----|
| 105 | 137 | 217 | 277 | 322 |

Average Turnaround time = (105 + (217-2) + (277-4) + (322-8) + (137-16))/5 = 1028/5 = 205.6
Average Waiting time = (0 + (137-2) + (217-4) + (277-8) + (105-16))/5 = 706/5 = 141.2

**CSPDABRR:**

Table 12: The scheduling table of CSPDABRR for CASE-V in uniprocessor environment.

| Round | 1st | 2nd | | | | 3rd | | 4th |
|-------|-----|-----|---|---|---|-----|---|-----|
| Process | P1 | P2 | P3 | P4 | P5 | P2 | P3 | P2 |
| B.T | 105 | 80 | 60 | 45 | 32 | 26 | 6 | 10 |
| A.T | 0 | 2 | 4 | 8 | 16 | 2 | 4 | 2 |
| PFP1 | 2 | 1 | 2 | 1 | 1 | 1 | 2 | 1 |
| PFP2 | 2 | 2 | 3 | 3 | 2 | 2 | 3 | 2 |
| PFP3 | 2 | 1 | 2 | 1 | 2 | 1 | 2 | 1 |
| PFP4 | 3 | 3 | 3 | 3 | 3 | 1 | 1 | 1 |
| PFP5 | 1 | 2 | 1 | 1 | 1 | 2 | 1 | 2 |
| PFP6 | 4 | 4 | 4 | 3 | 2 | 4 | 4 | 4 |
| PFP7 | 1 | 2 | 2 | 1 | 1 | 2 | 1 | 1 |
| Total | 15 | 15 | 17 | 13 | 12 | 13 | 14 | 12 |
| P.L | B.N | B.N | L | N | A.N | N | N | A.N |

| ←105→ | ← | 54 | → | ← | 16 | → | ← 10 → |
|-------|---|----|---|---|----|---|--------|
| P1 | P5 | P4 | P2 | P3 | P2 | P3 | P2 |
| 105 | 137 | 182 | 236 | 290 | 306 | 312 | 322 |

Average Turnaround time = (105 + (322-2) + (312-4) + (182-8) + (137-16))/5 = 1028/5 = 205.6
Average Waiting time = (0 + (242-2) + (252-4) + (137-8) + (105-16))/5 = 706/5 = 141.2

b. Multiprocessor Environment: (with 2 processors)

**RR(Round Robin):**
Average Turnaround time = (174 + (150-2) + (144-4) + (120-8) + (109-16))/5 = 667/5 = 133.4
Average Waiting time = (69 + (70-2) + (84-4) + (75-8) + (77-16))/5 = 345/5 = 69

**Priority:**
Average Turnaround time = (105 + (82-2) + (165-4) + (159-8) + (114-16))/5 = 595/5 = 119
Average Waiting time = (0 + (2 - 2) + (105 - 4) + (114 - 8) + (82 - 16))/5 = 273/5 = 54.6

**CSPDABRR:**

Table 13: The scheduling table of CSPDABRR for CASE-V in multiprocessor environment.

| Round   | 1$^{st}$ | 2$^{nd}$ |     | 3$^{rd}$ |     | 4$^{th}$ |
|---------|----|----|----|----|----|----|
| Process | P1 | P2 | P3 | P4 | P5 | P3 |
| B.T     | 105| 80 | 60 | 45 | 32 | 15 |
| A.T     | 0  | 2  | 4  | 8  | 16 | 4  |
| PFP1    | 2  | 1  | 2  | 1  | 1  | 2  |
| PFP2    | 2  | 2  | 3  | 3  | 2  | 3  |
| PFP3    | 2  | 1  | 2  | 1  | 2  | 2  |
| PFP4    | 3  | 3  | 3  | 3  | 3  | 1  |
| PFP5    | 1  | 2  | 1  | 1  | 1  | 1  |
| PFP6    | 4  | 4  | 4  | 3  | 2  | 4  |
| PFP7    | 1  | 1  | 2  | 1  | 1  | 1  |
| Total   | 15 | 14 | 17 | 13 | 12 | 14 |
| P.L     | B.N| N  | L  | N  | AN | N  |

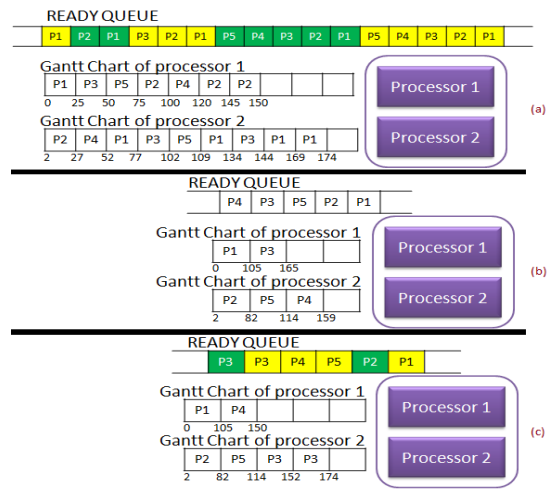

Figure 7: Pictorial representation of Round Robin, Priority, and CSPDABRR for CASE-V in Multiprocessor environment

Average Turnaround time = (105 + (82-2) + (174-4) + (150-8) + (114-16)/5 = 595/5 = 119
Average Waiting time = (0 + (2-2) + (114-4) + (105-8) + (82-16))/5 = 273/5 = 54.6

CASE6.          Processes with arrival time and with Random Burst time

a. Uniprocessor Environment:

Table 14: Process of CASE-VI.

| Process | P1 | P2 | P3 | P4 | P5 |
|---------|----|----|----|----|----|
| A.T     | 0  | 5  | 8  | 15 | 20 |
| B.T     | 45 | 90 | 70 | 38 | 55 |
| Priority| 5  | 1  | 3  | 4  | 2  |

**RR(Round Robin):**

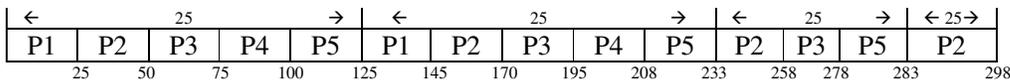

Average Turnaround time = (145+ (298-5) + (278-8) + (208-15) + (283-20))/5 = 1164/5 = 232.8
Average Waiting time = (100 + (208-5) + (208-8) + (170-15) + (228-20))/5 = 866/5 = 173.2

**Priority:**

| P1 | P2 | P5 | P3 | P4 |
|---|---|---|---|---|
| 45 | 135 | 190 | 260 | 298 |

Average Turnaround time = (45 + (135-5) + (260-8) + (298-15) + (190-20))/5 = 880/5 = 176
Average Waiting time = (0 + (45-5) + (190-8) + (260-15) + (135-20))/5 = 582/5 = 116.4

**CSPDABRR:**

Table 15: The scheduling table of CSPDABRR for CASE-VI in uniprocessor environment.

| Round | 1st | 2nd | | | | 3rd | | 4th |
|---|---|---|---|---|---|---|---|---|
| Process | P1 | P2 | P3 | P4 | P5 | P2 | P3 | P2 |
| B.T | 45 | 90 | 70 | 38 | 55 | 27 | 7 | 10 |
| A.T | 0 | 5 | 8 | 15 | 20 | 5 | 8 | 5 |
| PFP1 | 1 | 2 | 2 | 1 | 1 | 2 | 2 | 2 |
| PFP2 | 2 | 3 | 2 | 3 | 2 | 3 | 2 | 3 |
| PFP3 | 3 | 3 | 1 | 1 | 3 | 3 | 1 | 3 |
| PFP4 | 3 | 3 | 3 | 3 | 3 | 1 | 1 | 1 |
| PFP5 | 2 | 2 | 1 | 1 | 2 | 2 | 1 | 2 |
| PFP6 | 4 | 3 | 4 | 3 | 2 | 3 | 4 | 3 |
| PFP7 | 1 | 2 | 2 | 1 | 1 | 2 | 1 | 1 |
| Total | 16 | 18 | 15 | 13 | 14 | 16 | 12 | 15 |
| P.L | B.N | L | B.N | N | N | B.N | A.N | B.N |

| ← 45 → | ← | 63 | | → | ← 17 → | | ← 10 → |
|---|---|---|---|---|---|---|---|
| P1 | P4 | P5 | P3 | P2 | P3 | P2 | P2 |
| 45 | 83 | 138 | 201 | 264 | 271 | 288 | 298 |

Average Turnaround time = (45 + (298-5) + (271-8) + (83-15) + (138-20))/5 = 787/5 = 157.4
Average Waiting time = (0 + (208-5) + (201-8) + (45-15) + (83-20))/5 = 489/5 = 97.8

b. Multiprocessor Environment: (with 2 processors)

**RR(Round Robin):**

Average Turnaround time
= (75 + (158-5) + (145-8) + (113-15) + (143-20)/5
= 586/5 = 117.2
Average Waiting time
= (30 + (68-5) + (75-8) + (75-15) + (88-20))/5
= 288/5 = 57.6

**Priority:**

Average Turnaround time
= (45 + (95-5) + (165-8) + (138-15) + (100-20))/5
= 495/5 = 99
Average Waiting time
= (0 + (5-5) + (95-8) + (100-15) + (45-20))/5
= 197/5 = 39.4

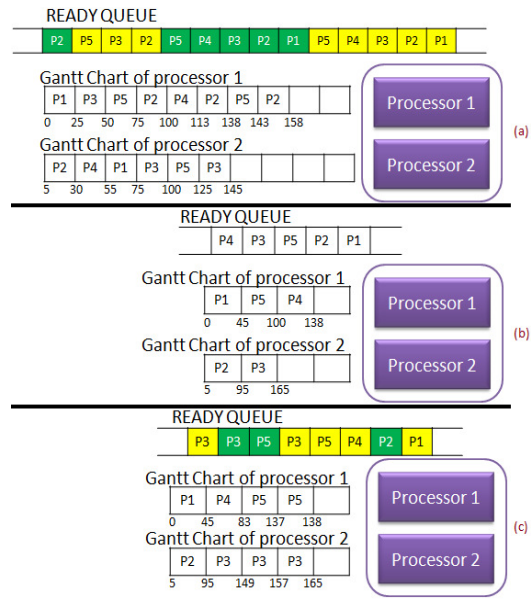

Figure 8: Pictorial representation of Round Robin, Priority, and CSPDABRR for CASE-VI in Multiprocessor environment

**CSPDABRR:**

Table 16: The scheduling table of CSPDABRR for CASE-VI in multiprocessor environment.

| Round   | 1st | 2nd | 3rd |    |    | 4th |    | 5th |
|---------|-----|-----|-----|----|----|-----|----|-----|
| Process | P1  | P2  | P3  | P4 | P5 | P3  | P5 | P3  |
| B.T     | 45  | 90  | 70  | 38 | 55 | 16  | 1  | 8   |
| A.T     | 0   | 5   | 8   | 15 | 20 | 8   | 20 | 8   |
| PFP1    | 1   | 2   | 2   | 1  | 1  | 2   | 1  | 2   |
| PFP2    | 2   | 3   | 2   | 3  | 2  | 2   | 2  | 2   |
| PFP3    | 3   | 3   | 1   | 1  | 3  | 1   | 3  | 1   |
| PFP4    | 3   | 3   | 3   | 3  | 3  | 1   | 1  | 1   |
| PFP5    | 2   | 2   | 1   | 1  | 2  | 1   | 2  | 1   |
| PFP6    | 4   | 3   | 4   | 3  | 2  | 4   | 2  | 4   |
| PFP7    | 1   | 1   | 2   | 1  | 2  | 2   | 1  | 1   |
| Total   | 16  | 17  | 15  | 13 | 15 | 13  | 12 | 12  |
| P.L     | B.N | L   | B.N | N  | B.N| N   | A.N| A.N |

Average Turnaround time = ((45-0) + (95-5) + (165-8) + (83-15) + (138-20))/5 = 478/5 = 95.6
Average Waiting time = (0 + (5-5) + (95-8) + (45-15) + (83-20))/5 = 180/5 = 36

## 5. RESULT ANALYSIS

From the analysis of all the six cases, it is concluded that our algorithm performs better than Round Robin and priority in both of performance metics and priority features. First coming to performance metics, the table and fig depicts the waiting time and turn-around time of round robin scheduling algorithm, priority scheduling algorithm, and CSPDABRR scheduling algorithm in case of uni-processor system. As per analysis the proposed algorithm saves 25.5% of turn-around time & 35.2% of waiting time than round robin scheduling algorithm in uni-processor environment. The proposed algorithm saves 12.8% of turn-around time & 18.9% of waiting time than priority scheduling algorithm in uniprocessor environment.

Table 17: analysis of turnaround time and waiting time obtained using Round Robin, Priority, and CSPDABRR in Uniprocessor environment.

| | Uni-Processor Environment | | | | | | | | | | | | | |
|---|---|---|---|---|---|---|---|---|---|---|---|---|---|---|
| | Waiting Time | | | | | | | Turnaround Time | | | | | | |
| | Case-I | Case-II | Case-III | Case-IV | Case-V | Case-VI | Total | Case-I | Case-II | Case-III | Case-IV | Case-V | Case-VI | Total |
| Round Robin | 192.00 | 144.00 | 62.60 | 144.40 | 204.20 | 173.20 | **920.40** | 261.40 | 196.60 | 99.20 | 205.60 | 268.60 | 232.80 | **1264.20** |
| Priority | 151.60 | 143.80 | 55.60 | 126.60 | 141.20 | 116.40 | **735.20** | 221.00 | 196.40 | 93.00 | 187.80 | 205.60 | 176.00 | **1079.80** |
| CSPDABRR | 134.60 | 69.00 | 51.60 | 102.00 | 141.20 | 97.80 | **596.20** | 204.00 | 121.60 | 89.00 | 163.20 | 205.60 | 157.40 | **940.80** |
| **Total** | **478.20** | **356.80** | **169.80** | **373.00** | **486.60** | **387.40** | **2251.80** | **686.40** | **514.60** | **281.20** | **556.60** | **679.80** | **566.20** | **3284.80** |

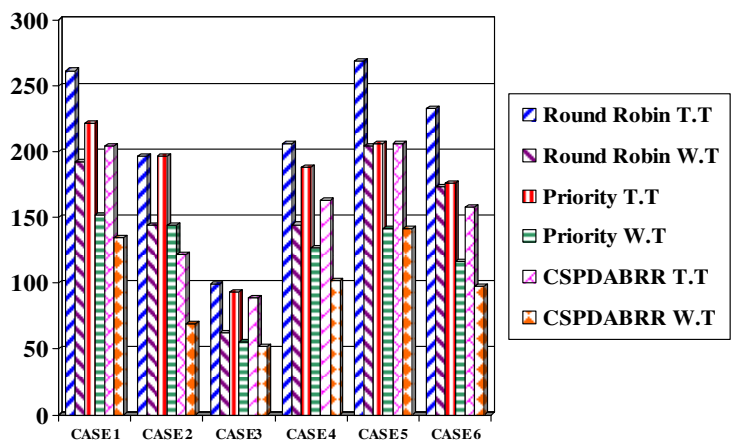

Figure 9: Comparative Performance analysis of Round Robin, Priority, and CSPDABRR in Uniprocessor environment.

Table 18: analysis of turnaround time and waiting time obtained using Round Robin, Priority, and CSPDABRR in Multiprocessor environment.

| | Multi-Processor Environment | | | | | | | | | | | | | |
|---|---|---|---|---|---|---|---|---|---|---|---|---|---|---|
| | Waiting Time | | | | | | | Turnaround Time | | | | | | |
| | Case-I | Case-II | Case-III | Case-IV | Case-V | Case-VI | Total | Case-I | Case-II | Case-III | Case-IV | Case-V | Case-VI | Total |
| Round Robin | 63.00 | 47.80 | 16.00 | 40.20 | 69.00 | 57.60 | **293.60** | 132.40 | 100.40 | 53.40 | 101.40 | 133.40 | 117.20 | **638.20** |
| Priority | 52.40 | 56.60 | 16.40 | 31.60 | 54.60 | 39.40 | **251.00** | 121.80 | 109.20 | 53.80 | 92.80 | 119.00 | 99.00 | **595.60** |
| CSPDABRR | 42.00 | 17.80 | 12.40 | 26.20 | 54.60 | 36.00 | **189.00** | 111.40 | 70.40 | 49.50 | 87.40 | 119.00 | 95.60 | **533.30** |
| **Total** | 157.40 | 122.20 | 44.80 | 98.00 | 178.20 | 133.00 | **733.60** | 365.60 | 280.00 | 156.70 | 281.60 | 371.40 | 311.80 | **1767.10** |

The table and fig depicts the waiting time and turn-around time of round robin scheduling algorithm, priority scheduling algorithm, and CSPDABRR scheduling algorithm in case of multi-processor system. As per analysis the proposed algorithm saves 16.4% of turn-around time & 35.6% of waiting time than round robin scheduling algorithm in multi-processor environment. The proposed algorithm saves 10.4% of turn-around time & 24.7% of waiting time than priority scheduling algorithm in multi-processor environment

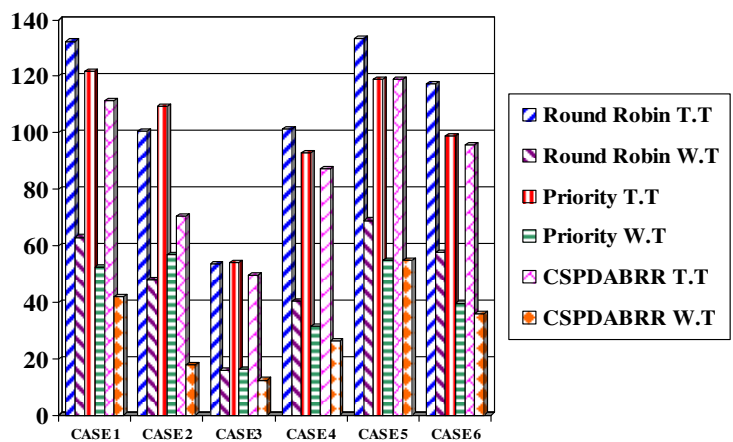

Figure 10: Comparative Performance analysis of Round Robin, Priority, and CSPDABRR in Multiprocessor environment

In case of priority, due to inclusion of these seven priority feature points it helps in proper management of processes according to their features. Our algorithm also helps in load balancing, deadlock avoidance, and proper utilization of process.

## 6. CONCLUSIONS

This paper presents the features of CSPDABRR algorithm. This algorithm is integrated of PFPs for better organization of processes according to their characteristics, features of round robin for better performance metics. Comparative analysis of RR scheduling algorithm, priority scheduling algorithm, and the proposed algorithm CSPDABRR has been carried out in both uni-processor and multi processor environment. The proposed algorithm provides better performance metrics by minimizing the average waiting time and average turnaround time. it also provide proper load balancing and proper arrangement of processes according to their features. In future we want to improve this algorithm by including the concept of multi-threading and processor affinity.

## REFERENCES

[1] Abraham Silberschatz, Peter B. Galvin, Greg Gagne (2009) "Operating System Concepts", eighth edition, Wiley India.
[2] Neetu Goel, R.B. Garg, (2012) "A Comparative Study of CPU Scheduling Algorithms", International Journal of Graphics and Image Processing Volume 2 issue 4, pp 245-251.
[3] Jayashree S. Somani, Pooja K. Chhatwani, (2013) "Comparative Study of Different CPU Scheduling Algorithms", International Journal of Computer Science and Mobile Computing, PP 310-318.
[4] Rami J. Matarneh (2009) "Self-Adjustment Time Quantum in Round Robin Algorithm Depending on Burst Time of the Now Running Processes", American Journal of Applied Sciences, pp 1831-1837.
[5] Abbas Noon, Ali Kalakech, Seifedine Kadry, (2011) "A New Round Robin Based Scheduling Algorithm for Operating Systems: Dynamic Quantum Using the Mean Average", International Journal of Computer Science Issues (IJCSI), Vol. 8(3), pp 224-229.
[6] H.S.Behera, R. Mohanty, Debashree Nayak, (2010) "A New Proposed Dynamic Quantum with Re-Adjusted Round Robin Scheduling Algorithm and Its Performance", International Journal of Computer Applications, pp 10-15.
[7] Ajit Singh, Priyanka Goyal, Sahil Batra (2010) "An Optimized Round Robin Scheduling Algorithm for CPU Scheduling", International journal on Computer Science and Engineering, pp 2383-2385.
[8] Manish Kumar Mishra, Dr. Faizur Rashid (2014) "An Improved Round Robin CPU Scheduling Algorithm with Varying Time Quantum", International Journal of Computer Science, Engineering and Applications (IJCSEA), pp 1-8.
[9] Rishi Verma, Sunny Mittal, Vikram Singh (2014) "A Round Robin Algorithm using Mode Dispersion for Effective Measure", International Journal for Research in Applied Science and Engineering Technology (IJRASET), pp 166-174.
[10] Radhe Shyam, Sunil Kumar Nandal, (2014) "Improved Mean Round Robin with Shortest Job First Scheduling", International Journal of Advanced Research in Computer Science and Software Engineering, PP 170-179.
[11] Amar Ranjan Dash, Sandipta kumar Sahu, Sanjay Kumar Samantra (2015) "An Optimized Round Robin CPU Scheduling Algorithm with Dynamic Time Quantum", International Journal of Computer Science, Engineering and Information Technology (IJCSEIT), Vol. 5,No.1, pp7-26.
[12] Bashir Alam, M.N. Doja, R. Biswas, M. Alam, (2011) "Fuzzy Priority CPU Scheduling Algorithm", International Journal of Computer Science Issues (IJCSI), Vol. 8(6), pp 386-390.
[13] H. S. Behera, Ratikanta Pattanayak, Priyabrata Mallick, (2012) "An Improved Fuzzy-Based CPU Scheduling (IFCS) Algorithm for Real Time Systems", International Journal of Soft Computing and Engineering (IJSCE), Vol. 2(1), pp 326-330.
[14] M.Ramakrishna, G.Pattabhi Rama Rao, (2013) "EFFICIENT ROUND ROBIN CPU SCHEDULING ALGORITHM FOR OPERATING SYSTEMS", International Journal of Innovative Technology and Research (IJITR), Vol. 1(1), pp 103-109
[15] Ishwari Singh Rajput, Deepa Gupta, (2012) "A Priority based Round Robin CPU Scheduling Algorithm for Real Time Systems", International Journal of Innovations in Engineering and Technology (IJIET), Vol. 1(2), pp pp 1-11.

[16] H.S.Behera, Sabyasachi Sahu, Sourav Kumar Bhoi, (2011) "Weighted Mean Priority Based Scheduling for Interactive Systems", Journal of Global Research in Computer Science, Vol. 2(5), pp 1-7.

[17] H.S. Behera, Simpi Patel, Bijayalakshmi Panda, (2011) "A New Dynamic Round Robin and SRTN Algorithm with Variable Original Time Slice and Intelligent Time Slice for Soft Real Time Systems", International Journal of Computer Applications, Vol. 16(1), PP.

[18] Rakesh Mohanty, H. S. Behera, Khusbu Patwari, Monisha Dash, M. Lakshmi Prasanna, (2011) "Priority Based Dynamic Round Robin (PBDRR) Algorithm with Intelligent Time Slice for Soft Real Time Systems", International Journal of Advanced Computer Science and Applications (IJACSA), Vol 2(2), PP 46-50.

[19] Zena Hussain Khalil, Ameer Basim Abdulameer Alaasam, (2013) "Priority Based Dynamic Round Robin with Intelligent Time Slice and Highest Response Ratio Next Algorithm for Soft Real Time System", Global Journal of Advanced Engineering Technologies, Vol 2(3), pp 120-124.

[20] H.S. Behera, Jajnaseni Panda, Dipanwita Thakur, Subasini Sahoo, (2011) "A New Proposed Two Processor Based CPU Scheduling Algorithm with Varying Time quantum for Real Time Systems", Journal of Global Research in Computer Science, Vol. 2(4), pp 81-87.

## AUTHORS

Amar Ranjan Dash has achieved his B. Tech. degree from Biju Patnaik University of Technology, Odisha, India and M. Tech. degree from Berhampur University, Odisha, India. His research interests include CPU Scheduling, Web Accessibility, and Cloud Computing. 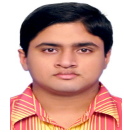

Sandipta Kumar Sahu has achieved his B. Tech. degree from Biju Patnaik University of Technology, Odisha, India and M. Tech. degree in computer science and engineering at National Institute of Science And Technology, Odisha, India. His research interests include Operating System, Software engineering, and Computer Architecture. 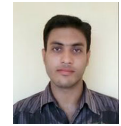

Sanjay Kumar Samantra has achieved his MCA degree from Berhampur University, Odisha, India and M. Tech. degree in computer science and engineering at National Institute of Science And Technology, Odisha, India. His research interests include CPU scheduling, Grid computing, and Cloud Computing. 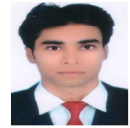

Sradhanjali Sabat has achieved her B.Tech degree from The Techno School, Odisha, India and M. Tech. degree in computer science and engineering at National Institute of Science And Technology, Odisha, India. Her research interests include wireless sensor networks, computer networks. 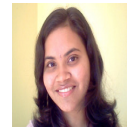